\documentclass{aa}
\usepackage{graphicx}

\begin{document}

\title{Rotation curve bifurcations as indicators of close recent galaxy encounters}

\author{S. Pedrosa \inst{1,2}
        \and P. B. Tissera \inst{1,2}
        \and I. Fuentes-Carrera \inst{3}
        \and C. Mendes de Oliveira \inst{4}
        }

\offprints{S. Pedrosa}

\institute{Consejo Nacional de Investigaciones Cient\'{\i}ficas y T\'ecnicas, CONICET, Argentina\\
\email{supe@iafe.uba.ar}
\and Instituto de Astronom\'{\i}a y F\'{\i}sica del Espacio, Casilla de Correos 67, Suc. 28, 1428, Buenos Aires, Argentina\\
\email{patricia@iafe.uba.ar}
\and GEPI, Observatoire de Paris, CNRS, Universit\'e Paris Diderot, Place Jules Janssen 92190 Meudon, France\\
\email{isaura.fuentes@obspm.fr}
\and Universidade de S\~ao Paulo, Instituto de Astronomia, Geof\'{\i}sica e Ci\^encias Atmosf\'ericas, Departamento de Astronomia, Rua do Mat\~ao 1226 - Cidade Universit\'aria 05508-900 S\~ao Paulo SP - Brazil\\
\email{oliveira@astro.iag.usp.br}
}

\date{Received / Accepted}

\abstract {Rotation curves of interacting galaxies often show that
velocities are either rising or falling in the direction of the
companion galaxy.} {We seek to reproduce and analyse these features
in the rotation curves of simulated equal-mass galaxies suffering a
one-to-one encounter, as possible indicators of close encounters.}
{Using simulations of  major mergers in 3D, we study the time
evolution of these asymmetries in a pair of galaxies, during the
first passage.} {Our main results are: (a) the rotation curve
asymmetries appear right  at pericentre of the first passage, (b)
the significant disturbed rotation velocities occur within a small
time interval, of $\sim 0.5 \rm Gyr  \ h^{-1}$, and therefore the presence of bifurcation
in the velocity curve could be used as an indicator of the
pericentre occurrence. These results are in qualitative agreement
with previous findings for minor mergers and fly-byes.} {}

\keywords{galaxies: interactions -- galaxies: kinematics and dynamics -- galaxies: spiral}

\titlerunning{Galaxy interactions and rotation velocity}
\authorrunning{Pedrosa et al.}

\maketitle

\section{Introduction}

\begin{figure}
\hspace{1cm}\resizebox{7cm}{!}{\includegraphics[scale=0.01]{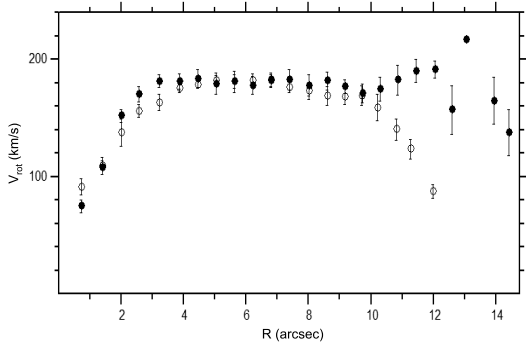}}
\hspace{-0.2cm} \caption{Rotation curve of the spiral galaxy NGC
5953 in the interacting galaxy pair KPG 468 (\cite {Her03}) within
the inner 15", derived from a full 2D velocity map. A velocity
bifurcation between the approaching side (full circles) and receding
one (empty circles) can be seen at 9". The companion NGC 5954 is a
galaxy of similar size and morphological type} \label{RCobs}
\end{figure}

Rotation curves of interacting and merging galaxies are often highly
disturbed. First observations of \cite {Rub91} showed several cases
of sinusoidal rotation curves and asymmetries for a large fraction
of compact group galaxies obtained from long-slit spectroscopy.
Although \cite {Men03} have shown that many of these peculiarities
are smoothed when rotation curves (hereafter RC) are derived from 2D
velocity maps, some disturbances still remain, indicating their true
nature as a global feature of the galaxy.
 For interacting galaxies, features on the RCs
are related not only to the structure of the disk itself, but also
to the external perturbations due to the companion (e.g. Rubin and
Ford 1983; Elmegreen and Elmegreen 1990; Chengalur et al. 1994; Salo
andLaurikainen 2000; Marquez et al. 2002; Hernandez Toledo et al.
2003; Fuentes Carrera et al. 2004; Garrido et al. 2005). One
noticeable perturbation in a rotation curve occurs when the velocity
on one side decreases considerably while the other side remains
fairly constant. Commonly these features, so called bifurcations,
happen when there is a close companion galaxy. An example of this
behaviour can be seen in Fig. ~\ref{RCobs} obtained by Fuentes
Carrera et al. (2004), where the RC of a galaxy in a pair of
galaxies of similar sizes and masses is displayed.

The natural question to ask here is how these velocity disturbances
relate to the passage of the companion galaxy and if they can be
used as a timer of the stage of  evolution of an interaction.
Recently, \cite {Kron06} (hereafter Kron06) have investigated
disturbed RCs in simulated interacting galaxies. Using simulations,
those authors reproduced the observations of intermediate redshift
galaxies (z$\sim$0.5) by putting a virtual slit along the major axis
of each investigated system (see also \cite  {Ba99}).
 They also found that the features associated with the perturbations in
the RCs strongly depend on the viewing angles, particularly, for the
derived asymmetry parameters (\cite{Da01}), which presented quite a
large dispersion, for minor mergers and fly byes. For major mergers,
no similar detailed study of the temporal behaviour of these
asymmetries has been made. This motivated us to revisit Kronberger's
study focusing on the analysis of the time evolution of the
asymmetries for {\it major mergers}. Furthermore, we chose to
peforme our analysis in 3D, to uncover the real level of disturbance
of the systems. For this purpose, we used N-body/hydrodynamic
simulations with cooling, star formation and supernova feedback.

\section{The Numerical Experiments}

We ran pre-prepared initial conditions  of  major mergers involving
Milky-Way type galaxies. We used the procedure developed by \cite
{Sprin05a} to construct near-equilibrium galaxies, each one
consisting of a dark matter halo, a disk of gas and stars and a
bulge, with a virial  velocity of $160 {\ \rm km \ s^{-1}}$. The
gaseous and stellar disks have exponential surface densities with an
scale-length, $ h_{d}$, of $2.5 {\ \rm kpc \ h^{-1}}$ and
scale-height of $ 0.2 \ h_{d}$. The bulge component follows a
Hernquist profile (\cite {Hern90}) with scale-length of $ 0.2 \
h_{d}$. The dark matter halo is consistent with a Hernquist profile
scaled to match the inner density distribution of dark matter haloes
found in cosmological simulations (\cite {Nav96}) with a
concentration parameter $ c=9 $. We use 20000, 10000, 50000 and
30000 particles to represent the stellar disk and bulge, the gas and
the dark matter components, respectively, of a given galaxy.
Baryonic masses range from $1.9 \times 10^5 {\rm h^{-1}}$ $M_{\sun}$
for particles on the disk to $1.3 \times 10^5 {\rm h^{-1}}$
$M_{\sun}$ in the bulge component, while the dark matter particles
have $3 \times 10^8 {\rm h^{-1}}$ $M_{\sun}$. We assume a $10\%$
fraction of baryons initially in the form of gas. A higher
resolution simulation with 600000 total particles was also run to
discard resolution effects.

 The two identical
galaxies were given an initial relative velocity of $200 {\ \rm km \
s^{-1}}$. The minimum separation, $r_{p}$, achieved after $ 1.3 \
{\rm Gyr} \ h^{-1}$, during the first passage is $30 {\ \rm kpc \
h^{-1}}$. Galaxies are set on parabolic orbits and will merge due to
dynamical friction. We investigated four possible configurations for
the encounter. We ordered the experiments using the orbital
parameters of \cite {Toom72} as $i_{1}=0$ and $i_{2}=0$ for A,
$i_{1}=0$ and $i_{2}=180$  for B , $i_{1}=45$, $ i_{2}=-45$ and
$\omega_{2}=90$ for C, $i_{1}=-45$, $i_{2}=45$ and $\omega_{2}=270$
for D. These initial configurations were chosen to study the
kinematical behaviour of the galaxies between the extreme cases of a
parallel encounter versus a perpendicular one. For experiment A we
also run two simulations changing the minimum separation at
pericentre: A1 and A2 with $r_{p}=15 {\ \rm kpc \ h^{-1}}$ and
$r_{p}=50 {\ \rm kpc \ h^{-1}}$, respectively.

The simulations were run by using the Tree-SPH Gadget-2 code (\cite
{Sprin05}). The cold gas can form new stars following the Kennicutt
law according to a stochastic algorithm (\cite {Scan05}). However,
since the initial gas component is a small fraction of the total
mass, the impact of new stars is not important for this analysis.The
systems experienced a number of close encounters until they finally
merged. In this work we focus only on the perturbation arising
during the first close passage.

\section{Simulated Rotation Curves}

In a disk structure supported by rotation and in equilibrium within
its potential well, the only velocity component should be a
tangential velocity ($v_{\phi}$) on the plane perpendicular to the
angular momentum of the system. In practice, this is not forcibly
the case. Motions of stars and gas in the disk can be modified by
the presence of structures in the disk or by external perturbations.
In this case, particular features on the RCs, such as bifurcations,
may appear.

At a given time during the interaction, the simulated RCs are
constructed by calculating  $v_{\phi}$ on the rotation plane
(defined as the plane perpendicular to the total angular momentum of
the system) as a function of the distance to the mass centre of the galaxy ($r$).
 For the purposes of analysing
possible asymmetries in the RCs, we considered $v_{\phi}$ within
four cones of solid angle of 60 degrees along two orthonormal
directions on the disk plane: $\alpha - \beta$, $\delta - \gamma$. We
note that results are insensitive to variations of these
angles. In fact, we have varied these angles between 10 and 60
degrees, finding no significant changes in the trends. By analysing
the velocities and angular momentum of the mass within these angles,
we can study the effects of interactions in different regions of the
disks and look for loss of symmetry in the velocity patterns.

We calculated the mean tangential velocity, $<v_{\phi}>$, estimated
in equally spaced radial bins of size 1.5 {\rm kpc $h^{-1}$}. We
also estimated the predicted velocity that particles on circular
orbits within a potential well dominated by dark matter should
statistically have as $V_{\rm cir}(r)= (\frac {G M_{\rm tot}(<r)}
{r})^{0.5}$ where $M_{\rm tot}(< r)$ is the total mass (baryons plus
dark matter) within $r$. For  systems in rotational equilibrium
within such a potential well, we should find that $<v_{\phi}> \sim
V_{\rm cir}$. Important departures from this relation will be
assumed to be indicative of a loss or a gain of angular momentum by
the action of an external force. Before the galaxies start
interacting, the disks are in rotational equilibrium within their
potential wells, regardless of the angular sector chosen to measure
$<v_{\phi}>$.

\begin{figure}
\resizebox{6cm}{!}{\includegraphics{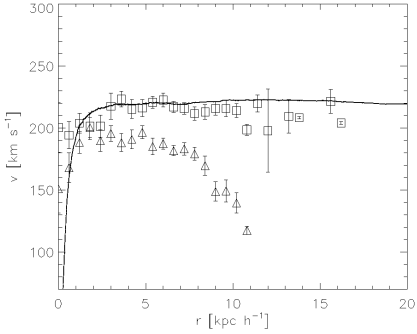}} \hspace{-0.2cm}
\caption{Circular (continuous line) and tangential velocities
(triangles and squares) in experiment A showing clear bifurcation.
Triangles and squares denote opposite orientations along a given
direction used to measure the curves. The side which shows a
declining tangential velocity is the one closest to the companion
galaxy. Error bars correspond to the standard deviations of the mean
values computed within each bin.} \label{velocidad}
\end{figure}

In order to assess the effects of the interaction on the rotation
curves, we analysed the departures of $<v_{\phi}>$ from $V_{\rm
cir}$ within each adopted angle as a function of time, for the four
experiments performed. Fig. ~\ref{velocidad} shows an example of a
disturbed RC in the $ \alpha - \beta $ direction
for experiment A. A very asymmetric response of the disk
can be seen. For this RC, the velocity on the left side of the
galaxy ($\beta$ direction) decreases considerably, while the
velocity on the right-side of the galaxy ($\alpha$ direction)
remains fairly constant, producing a bifurcation similar to those
reported by observations (Fig. ~\ref{RCobs}).

In order to study the presence of these bifurcations in the RCs
along the interaction event, we defined the residual velocities
between the RC and the circular velocity as: $R = < v_{\phi} > -
V_{\rm cir} $. By estimating $R$ as a function of radius, we can
quantify the difference - and its amplitude - between the measured
tangential velocity on one side of the galaxy and that on the
opposite side.

\begin{figure}
\resizebox{6cm}{!}{\includegraphics{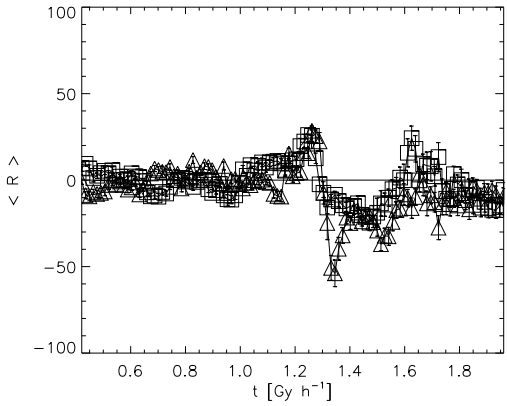}} \hspace{-0.2cm}
\caption{Velocity residuals  as a function of time estimated for
baryons within  $\alpha$ (squares) and $\beta$ (triangles) angles
 for experiment A.} \label{residuos}
\end{figure}

Since we are interested in the evolution of these differences with
time, we defined a mean residual velocity $<R>$ at each given time,
averaging out the corresponding residuals (the central regions of
the galaxies are excluded to avoid an artificial weakening of the
signal originated by the disk structure). Fig.~\ref{residuos} shows
the mean residuals calculated for experiment A as a function of time
in the $\alpha - \beta$ direction. From this figure, we can observe
that, in general, significant residuals are present, starting at $t
\approx 1.25 \ \rm  Gyr \ h^{-1}$, which corresponds to the time of peripassage.
We
note that the largest bifurcations appear when the RC is measured
within the angular sector closer to the companion galaxy. This can
be seen in  Fig. ~\ref{residuos} where the $\beta$ direction, which
is the one closest to the companion, shows the largest mean
residuals. A detailed analysis of the angular momentum content of
the simulated galaxies shows that these residuals can be directly
linked to the variation of angular momentum on orthogonal
directions.

\section{Can bifurcations be used as a timer for recent interactions?}

\begin{figure*}
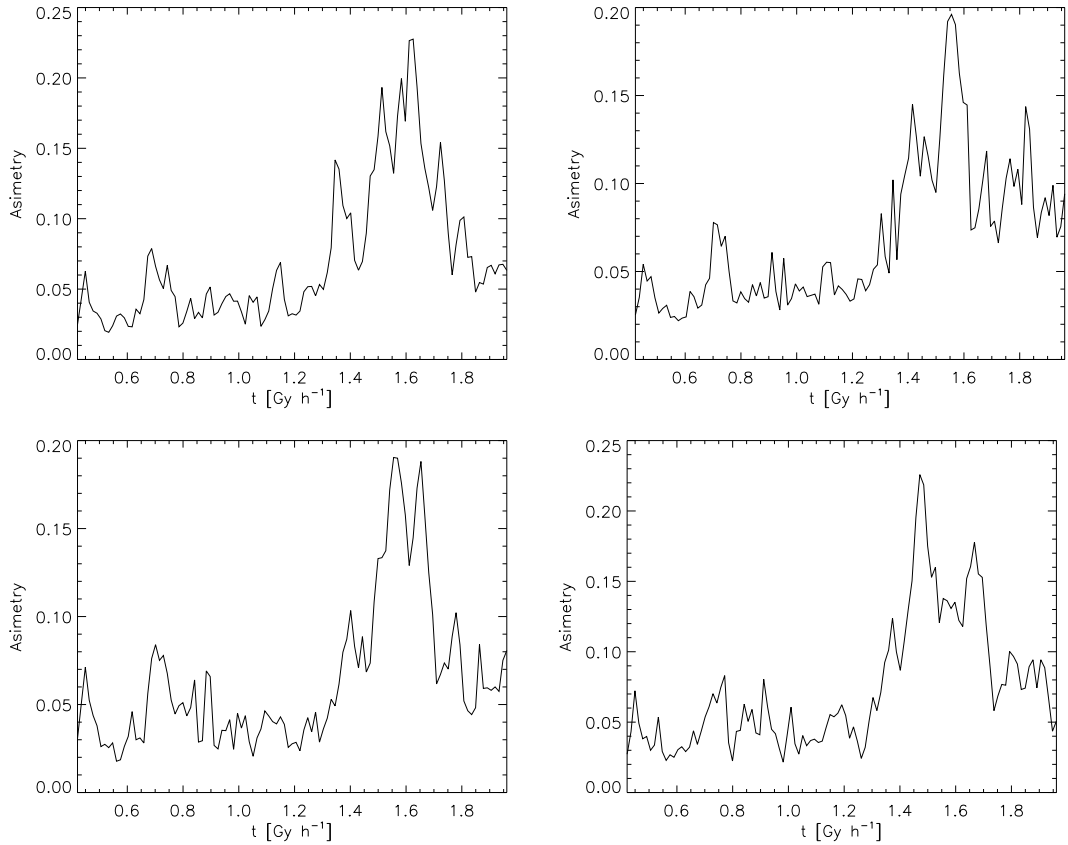

 \hspace*{1.8cm}\resizebox{6cm}{!}{\includegraphics{Figure3c.epsi}}
 \hspace*{1.2cm}\resizebox{6cm}{!}{\includegraphics{Figure3d.epsi}} \hspace*{-0.2cm}
\caption{Asymmetry parameter as a function of time for our four
experiments (A, B, C, D) along the $\alpha - \beta$ directions (see
Section 2 for experiment description).} \label{Asy}
\end{figure*}

In order to quantify if bifurcations in the RCs can be used as a
timer of the stage in the interaction, we analyse the evolution with
time of the global RC asymmetry, defined by Dale et al. (2001). This
asymmetry parameter (hereafter $AP$) is calculated as the area
between the approaching and receding halves of the RC, normalized by
the average area under the curve. Fig.~\ref{Asy} shows that when the
simulated galaxies have galaxy-galaxy distances of approximately $r
= 30 {\ \rm kpc \ h^{-1}}$, significant signal of asymmetry appears.
Similar levels of $AP$ are detected in our four experiments,
although the details of the features vary from one to the other.

Hence, according to the trends shown in Fig.~\ref{residuos} and Fig.~\ref{Asy},
two galaxies need to be very close
to start experiencing a significant, internal re-distribution of
angular momentum and mass, but when this has occurred, the RCs
remain perturbed for an extended period until they reach equilibrium
again, after almost $ 1 \ {\rm Gyr \  h^{-1}}$ of the pericentre,
when the centres of masses are separated by more than $ 200 \  {\rm
kpc  h^{-1}}$, in agreement with previous findings for minor mergers
and fly-byes (e.g. Kron06).

A detailed analysis of the evolution of the $AP$ displayed in
Fig.~\ref{Asy} shows that before the peripassage, for the four
experiments remains approximately at a value of $5\%$. From the
peripassage, a noticeable increase is detected with values up
$\approx$ 20\% and greater than 10\%, and with a mean $AP$ that, at
least,  doubles the level registered up to peripassage. After $ 0.5
\rm \ Gyr \  h^{-1}$ from peripassage, it starts decreasing. A
similar analysis for A1 and A2 shows that $AP$ anticorrelates with
$r_{p}$. We find no $AP$ greater than $10\%$ for $r_{p}$ larger than
$ \approx 50 {\ \rm kpc \ h^{-1}}$ in this merger configuration.

\begin{figure*}
\hspace*{1.8cm}\resizebox{6cm}{!}{\includegraphics{Figure3a.epsi}}
\hspace*{1.2cm} \resizebox{6cm}{!}{\includegraphics{Figure3b.epsi}}
\hspace*{-0.2cm}
\end{figure*}

We calculated the $AP$ for the RCs of 3 galaxies in isolated pairs
of galaxies of similar masses, studied by Fuentes-Carrera
(\cite{Her03}) and 2 galaxies in Hickson Compact Groups observed by
(Mendes de Oliveira et al. 2003) which show bifurcations. All RCs
were derived from full 2D velocity field maps by averaging large
velocity cones around the major axes of the galaxies, and therefore
they are good representations of their kinematics as a whole. They
have been corrected by inclination so that the comparison with
simulations is as good as it can be carried out. In the case of the
pairs studied by Fuentes Carrera the asymmetry values obtained are:
$19\%$ for NGC5426, $13\%$  NGC5427 and $9\%$ for NGC5953. A similar
behaviour is obtained for the galaxies in groups: $13\%$  for HCG7c
and $36\%$ for HCG96a. We have also calculated $AP$ for
non-bifurcated-RC galaxies (HCG 10d, HCG88a, HCG 88c, HCG 88d, HCG
89a, HCG 91c1, HCG 96d) in compact groups. The obtained mean value
was $7\%$, which is very similar, qualitatively, to the one found in
the simulations, during the unperturbed phase ($5\%$). It is
interesting to note that the galaxies with bifurcated $RC$ in
compact groups are spiral galaxies with only one companion of
similar mass (HCG 07a and HCG 96c, respectively), and several other
considerably less massive companions (1/3 of the mass of the first
ranked galaxy, at most).

Hence, from the analysis of these high quality observed RCs we get similar results to
those found in simulations for the asymmetry value of the bifurcated RCs.
\section{Conclusions}

We studied the RCs of two simulated Milky Way type galaxies during
their first encounter. We performed four simulations with different
orientations of the angular momentum direction of the systems. The
simulated RCs displayed bifurcations consistently with observational
findings. The evolution with time of the features were investigated.

Our main conclusions are:

\begin{enumerate}

\item The asymmetry parameter (defined by \cite {Da01}) shows a clear
increase at the peripassage, reaching
values larger than a factor of two the levels detected
for RCs in equilibrium. After $\approx 0.5 \ {\rm Gyr \  h^{-1}} $,
the asymmetry parameter begins to decrease again to lower values.
We found that observed RCs corrected by inclination yield a similar behaviour.

\item The most significant asymmetries are detected  at peripassage
when the systems are separated $\approx 30 \ {\rm kpc \ h^{-1}}$. In
our experiments we found that the perturbations in the RCs can be
detected even after almost $ 1 \ {\rm Gyr \ h^{-1}}$ of the
peripassage in agreement with previous works for minor
mergers and fly byes (Kron06)

\end{enumerate}

Therefore the asymmetry parameter could be used to roughly estimate
if the pericentre has occurred in the last  $\approx 0.5 \ {\rm Gyr \
h^{-1}} $. The question if these features can always be seen in RCs
of interacting galaxies might depend on the viewing angle (Kron06).
 However, the good agreement of the
observational and numerical results obtained in this paper
encourages us to carry out a more comprehensive comparison with
observations in a future paper.

\begin{acknowledgements}

IFC acknowledges the financial support of FAPESP grant No.
03/01625-2 and the Sixth Framework Program of the EU for a Marie
Curie Postdoctoral Fellowship. CMdO acknowledges support from the
Brazilian agencies FAPESP (projeto tem\'atico 01/07342-7), CNPq and
CAPES. PBT thanks Conicet and Foncyt for their support. This work
was partially supported by the European Union's ALFA-II programme
(LENAC) Numerical simulations were run on HOPE, Beowulf Cluster at
IAFE.

\end{acknowledgements}

\end{document}